\PassOptionsToPackage{table}{xcolor}
\documentclass[manuscript]{acmart}

\renewcommand\footnotetextcopyrightpermission[1]{} 
\usepackage{caption}

\usepackage{graphicx}
\usepackage{float} 
\usepackage{subfig}
\usepackage{subcaption}
\usepackage{overpic}
\usepackage{lipsum}
\AtBeginDocument{%
  }

\setcopyright{acmlicensed}
\copyrightyear{2018}
\acmYear{2018}
\acmDOI{XXXXXXX.XXXXXXX}

\acmJournal{POMACS}
\acmVolume{37}
\acmNumber{4}
\acmArticle{111}
\acmMonth{8}

\begin{document}

\title{Exploring the Effects of Traditional Chinese Medicine Scents on Mitigating Driving Fatigue}

\author{Nengyue Su}
 \affiliation{
  \department{School of Computer Science and Engineering}
  \institution{University of Electronic Science and Technology of China}
  \city{Chengdu}
  \country{China}
 }
\author{Liang Luo}
 \affiliation{
  \department{School of Computer Science and Engineering}
  \institution{University of Electronic Science and Technology of China}
  \city{Chengdu}
  \country{China}
 }
\author{Yu Gu}
 \affiliation{
  \department{School of Computer Science and Engineering}
  \institution{University of Electronic Science and Technology of China}
  \city{Chengdu}
  \country{China}
 }
\author{Fuji Ren}
 \affiliation{
  \department{School of Computer Science and Engineering}
  \institution{University of Electronic Science and Technology of China}
  \city{Chengdu}
  \country{China}
 }
\renewcommand{\shortauthors}{Nengyue Su, et al.}


\begin{abstract}
The rise of autonomous driving technology has led to concerns about inactivity-induced fatigue. This paper explores Traditional Chinese Medicine (TCM) scents for mitigating. Two human-involved studies have been conducted in a high-fidelity driving simulator. Study 1 maps six prevalent TCM scents onto the arousal/valence circumplex to select proper candidates, i.e., argy wormwood (with the highest arousal) and tangerine peel (with the highest valence). Study 2 tests both scents in an auto-driving course. Statistics show both scents can improve driver alertness and reaction-time, but should be used in different ways: argy wormwood is suitable for short-term use due to its higher intensity but poor acceptance, while tangerine peel is ideal for long-term use due to its higher likeness. These findings provide insights for in-car fatigue mitigation to enhance driver safety and well-being. However, issues such as scent longevity as for aromatherapy and automatic fatigue prediction remain unresolved.

\end{abstract}

\ccsdesc[500]{Human-centered computing~Empirical studies in HCI}

\keywords{Perception, Smell, Odor Stimulation, Driving fatigue, In-Car systems}


\maketitle

\section{Introduction}
\begin{figure}
    \centering
    \includegraphics[width=1\linewidth]{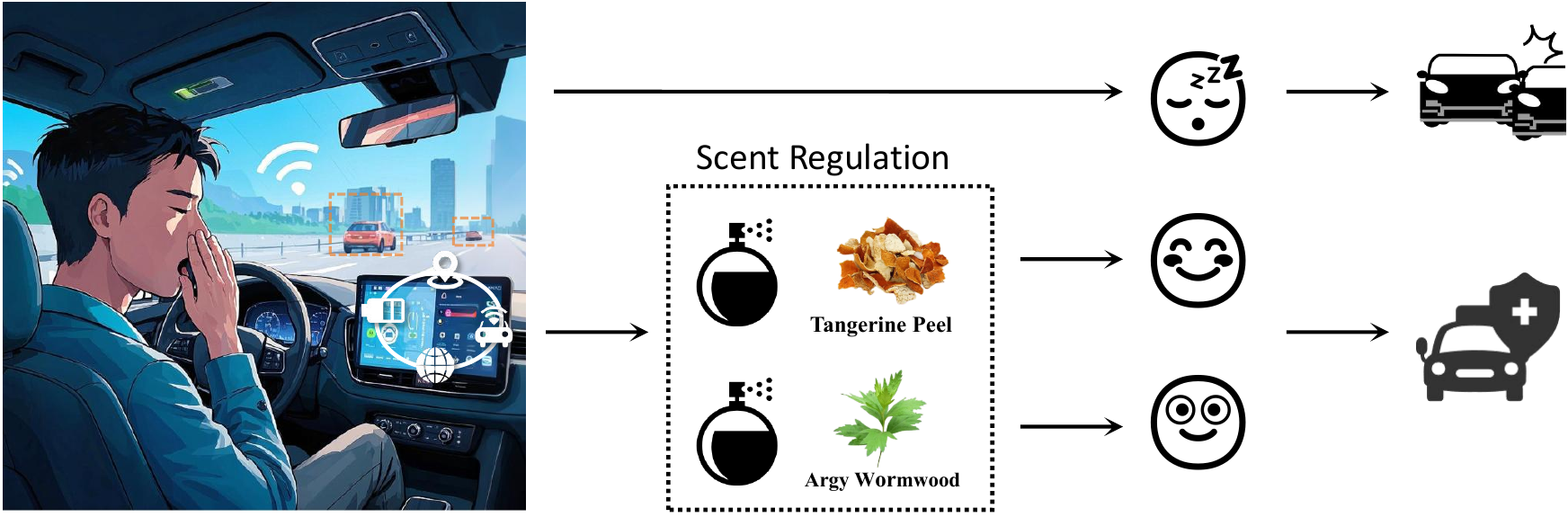}
    \caption{Our study aims to mitigate driving fatigue with TCM scents. As the left picture\protect\footnotemark[1] shows, autonomous driving can cause driving fatigue easily. Unregulated, it may lead to a car accident. Our research found that tangerine peel regulation can make drivers comfortable and alert, while argy wormwood regulation makes drivers less comfortable but more alert. Both benefit driving safety. }
    \label{fig:1}
\end{figure}

Currently, the emergence of autonomous driving technology has rendered driving more monotonous, thereby elevating the likelihood of passive fatigue (as in Figure~\ref{fig:1}) \cite{jamson2013behavioural}. This mental condition, marked by low valence and arousal, prolongs drivers' reaction-times \cite{eyben2010emotion} and, if left unaddressed, can lead to accidents \cite{li2024review}, posing substantial risks to road safety \cite{zhang2016traffic}.

Addressing driving fatigue often entails sensory stimuli, which can be categorized into visual \cite{li2021visual,qin2021characteristics,hassib2019detecting}, auditory \cite{orsini2024music,trumbo2017name}, and olfactory \cite{dahlman2024vehicle,jiang2023study} techniques. Considering the pivotal roles of vision and hearing in driving, olfactory modulation presents distinct advantages in maintaining driver functionality. Previous scent-related research has predominantly centered on Western essential oils, which may not resonate widely with Chinese populations. However, insights from Chinese aromatherapy traditions \cite{nan2013aromatherapy,giannenas2020history} hint that scents derived from Traditional Chinese Medicine (TCM) could be more culturally acceptable. Nevertheless, the influence of TCM scents on emotions and driving fatigue remains unexplored. Hence, this study endeavors to delve into this domain.

More specifically, two studies involving Chinese participants were conducted in a high-fidelity driving simulator. In Study 1, six TCM scents (see Fig. 2 for their appearances), i.e., acorus tatarinowii, argy wormwood, aromatic turmeric root-tuber, Chinese angelica, spine date seed, and tangerine peel, are mapped onto the arousal/valence circumplex(as in \cite{dmitrenko2020caroma}). Among them, argy wormwood (with the highest arousal) and tangerine peel (with the highest valence) were chosen for further examination . Study 2 aimed to explore the specific effects of these two scents on driving fatigue in a predesigned auto-driving course (prolonged driving on a straight highway with no traffic on a cloudy day). The results indicate that both scents can enhance driver alertness and reaction-time. However, their usage differs: argy wormwood is suitable for short-term applications due to its high intensity but lower acceptance, whereas tangerine peel is preferred for long-term use due to its higher popularity. 

The studies have examined the effect of Traditional Chinese Medicine (TCM) scents on mitigating driving fatigue, yet leaving unresolved questions such as the short duration of scent effectiveness in aromatherapy and the comparison between TCM and conventional flower and fruit scents. Furthermore, how to promptly detect driving fatigue? Additionally, could we develop a scent-based fatigue mitigation system that can predict the onset of driving fatigue by analyzing driving contexts (like the setting we used in the simulator) via Artificial Intelligence (AI)?

\footnotetext[1]{Generated by Doubao with prompt "Please generate a picture showing a side view of the inside of a car. In the picture, the driver is yawning with one hand covering their mouth and the other hand on their leg. There are autonomous driving icons like Wi-Fi and navigation on the steering wheel, and the car can automatically detect vehicles outside the window."}

\section{Related Work}
We structured the related work around two research areas:

\subsection{How to mitigate driving fatigue?}
Based on the classification of sensory stimuli, methods aimed at mitigating driving fatigue can be categorized into visual, auditory, and olfactory approaches. With regard to visual methods, adjustments to ambient lighting have been employed to enhance driver alertness \cite{li2021visual,qin2021characteristics,hassib2019detecting}. For auditory methods, music\cite{guo2024could,orsini2024music} or name music game\cite{trumbo2017name} have proven effective . In driving scenarios, vision and hearing serve as the primary channels for information reception. While modulation through these two sensory modalities may lead to driver distraction, olfaction is less utilized during driving and exhibits a closer relationship with emotional states\cite{alaoui1997basic}. Consequently, this study employs olfactory modulation as a means to address driving fatigue.

Contemporary research has substantiated that scents such as mint, grapefruit, lemon, and lavender can markedly ameliorate driving fatigue \cite{jiang2024scented,dahlman2024vehicle,yoshida2011study}. However, previous studies have predominantly focused on common flower and fruit scents that are widely utilized in Western countries. Given the disparities in culture and physiological traits, these scents are not frequently utilized among the Chinese populace \cite{yunjun2013perfume}. Consequently, there is a pressing need to explore scents that are more familiar and acceptable to Chinese individuals.

\subsection{Why we choose TCM?}
The Chinese populace has maintained a long-standing tradition of utilizing Traditional Chinese Medicine (TCM) to regulate emotions since ancient times \cite{farrar2020clinical}. In the framework of TCM, a bi-directional causal relationship exists between the harmonious flow of \textit{qi}\footnote[1]{Chinese identify \textit{qi} as the natural energy intrinsic to all things that exist in the universe, as the fundamental life energy responsible for health and vitality\cite{mccaffrey2003qigong}.} and both a healthy mental state and emotional equilibrium \cite{zhou2021conceptualization}. Specific TCM herbs, including Chinese angelica, acorus tatarinowii, spine date seed, and tangerine peel, have demonstrated efficacy in treating depression \cite{li2020traditional,zhang2019challenge,sun2022dissecting}.
However, TCM is predominantly administered for internal consumption. Relatively few TCM formulations have been processed into high-purity essential oils for aromatherapy purposes, and the precise effects of TCM scents on emotions remain unclear. Therefore, this study selects TCM essential oils with potential relevance to emotion regulation for investigation.

\section{Experiments}

\subsection{Overview of Experiments}
To explore the effects of TCM scents on mitigating driving fatigue, we designed two studies.
\begin{itemize}
    \item \textbf{Study 1:} Mapping TCM scents onto the arousal/valence circumplex. 
    \item \textbf{Study 2\protect\footnotemark[2]:} Exploring the effects of TCM scents on driving fatigue.
\end{itemize}
Both studies have passed the ethical review of the University of Electronic Science and Technology of China. All participants were carefully screened to make sure they had no breathing problems and no smell allergies.

\footnotetext[2]{To explore the duration of the driving fatigue induction experiment and he scent concentration of the scent release experiment in Study 2, a Pre-study was designed, as shown in the appendix. }

\subsection{Study 1: Mapping TCM scents onto the arousal/valence circumplex.}

\begin{figure}
    \begin{minipage}[]{0.48\textwidth}
        \vspace{0.0in}
        \centering
        \resizebox{1\linewidth}{!}{
            \begin{tabular}{|l|l|l|}
            \hline
            TCM                                                                                                                    & Ingredients                                                             & Efficacy                                                                                                                              \\ \hline
            \cellcolor[HTML]{F2F2F2}{\color[HTML]{08090C} Argy Wormwood}                                                           & \begin{tabular}[c]{@{}l@{}}Eucalyptol, \\ Caryophyllene\end{tabular}    & \begin{tabular}[c]{@{}l@{}}Warm the meridians to stop bleeding, \\ dispel cold and relieve pain, \\ calm anxiety\end{tabular}         \\ \hline
            \cellcolor[HTML]{F2F2F2}{\color[HTML]{08090C} Acorus Tatarinowii}                                                      & \begin{tabular}[c]{@{}l@{}}p - Cymene, \\ Nerol acetate\end{tabular}    & \begin{tabular}[c]{@{}l@{}}Open orifices and resolve phlegm, \\ remove dampness \\ and promote appetite\end{tabular}                  \\ \hline
            \cellcolor[HTML]{F2F2F2}{\color[HTML]{08090C} Chinese Angelica}                                                        & Ligustilide                                                             & \begin{tabular}[c]{@{}l@{}}Nourish and activate blood, \\ regulate mood\end{tabular}                                                  \\ \hline
            \cellcolor[HTML]{F2F2F2}{\color[HTML]{08090C} \begin{tabular}[c]{@{}l@{}}Aromatic Turmeric \\ Root-tuber\end{tabular}} & \begin{tabular}[c]{@{}l@{}}Camphene, Camphor,\\  Curcumene\end{tabular} & \begin{tabular}[c]{@{}l@{}}Promote blood circulation \\ to relieve pain, promote \textit{qi} circulation\\ and relieve depression\end{tabular} \\ \hline
            \cellcolor[HTML]{F2F2F2}{\color[HTML]{08090C} Spine Date Seed}                                                         & Suanzaoren fatty acids                                                  & \begin{tabular}[c]{@{}l@{}}Nourish \textit{qi} and \\ tonify the liver, calm the heart \\ and soothe the nerves\end{tabular}                   \\ \hline
            \cellcolor[HTML]{F2F2F2}{\color[HTML]{08090C} Tangerine Peel}                                                          & Hesperidin                                                              & \begin{tabular}[c]{@{}l@{}}Regulate \textit{qi} and \\ nourish the spleen, \\ dry dampness and resolve phlegm\end{tabular}                     \\ \hline
            \end{tabular}
        }
        \vspace{0.1in}
        \footnotetext{(a) Selected TCM}
    \end{minipage}
    \begin{minipage}[]{0.5\textwidth}
        \centering
        \includegraphics[width=1\linewidth]{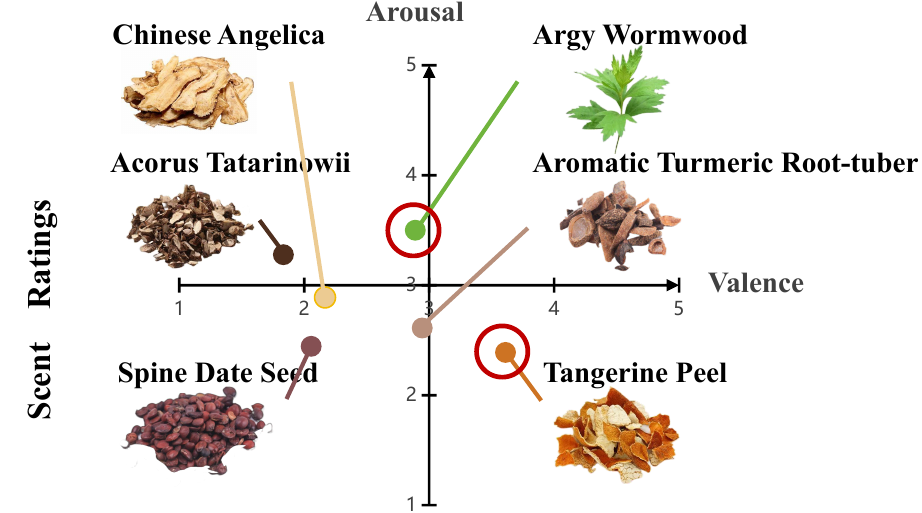}
        \footnotetext{(b) Arousal/valence circumplex}
    \end{minipage}
    \caption{Material and results of Study 1:(a) shows the ingredients and efficacy of the six TCM scents selected in Study 1. (b) presents the results of Study 1, mapping the TCM scents onto the arousal/valence circumplex.}
    \label{Fig.2}
\end{figure}

In this study, we conducted a screening of existing TCM essential oils and selected six that are specifically associated with emotional regulation for further research. Fig.\ref{Fig.2}(a) depicts the six TCM scents, along with their constituent ingredients and efficacy, as outlined in the Pharmacopoeia of the People's Republic of China \cite{2020Pharmacopoeia}.

\subsubsection{Design}
We conducted a within-participants study asking the participants to rate six scents (argy wormwood, acorus tatarinowii, Chinese angelica, aromatic turmeric root-tuber, spine date seed, and tangerine peel).

\subsubsection{Setup}
Participants sniffed the scents from identical bottles located on a table. Each bottle contained 10ml of essential oil. They sniffed each bottle for 2s, with intervals of 20s(as in \cite{wilson2017multi}).

\subsubsection{Procedure}
After sniffing each scent, participants were asked to rate the valence and arousal of each scent using Self-Assessment Manikins (SAM)\cite{bradley1994measuring} on a 5-point Likert scale (1 = "low", 5 = "high", see Fig.~\ref{Fig.2}(b)).

\subsubsection{Results}
18 participants, 23-26 years old (M = 24.5, SD = 1.04, 8 females), volunteered for this study. 
The normality test showed that the scent ratings were normally distributed. We conducted a repeated measures ANOVA test to analyze the data and found that scents had a main effect on both arousal (F = 2.509, p < 0.05) and valence (F = 9.823, p < 0.001). The results of this study (see Fig. ~\ref{Fig.2}(b)) confirmed that among the six TCM scents, argy wormwood (Arousal: M = 3.5, SD = 1.04; Valence: M = 2.89, SD = 0.76) and acorus tatarinowii (Arousal: M = 3.28, SD = 1.45; Valence: M = 1.83, SD = 0.92) had low valence and high arousal. Chinese angelica (Arousal: M = 2.89, SD = 1.18; Valence: M = 2.17, SD = 0.71), aromatic turmeric root-tuber (Arousal: M = 2.61, SD = 1.29; Valence: M = 2.84, SD = 0.99), and spine date seed (Arousal: M = 2.44, SD = 1.2; Valence: M = 2.06, SD = 0.99) had low valence and low arousal. tangerine peel (Arousal: M = 2.39, SD = 1.09; Valence: M = 3.61, SD = 1.04) had low arousal and high valence. Since arousal is an important indicator of driving fatigue\cite{jiang2024scented}, we chose argy wormwood which has the highest arousal values for Study 2. Considering its valence value is low, we chose tangerine peel which has a relatively high valence for comparison.

\begin{figure}[!t]
\centering
\subfloat[The location of the experimental equipment]{
		\includegraphics[scale=0.35]{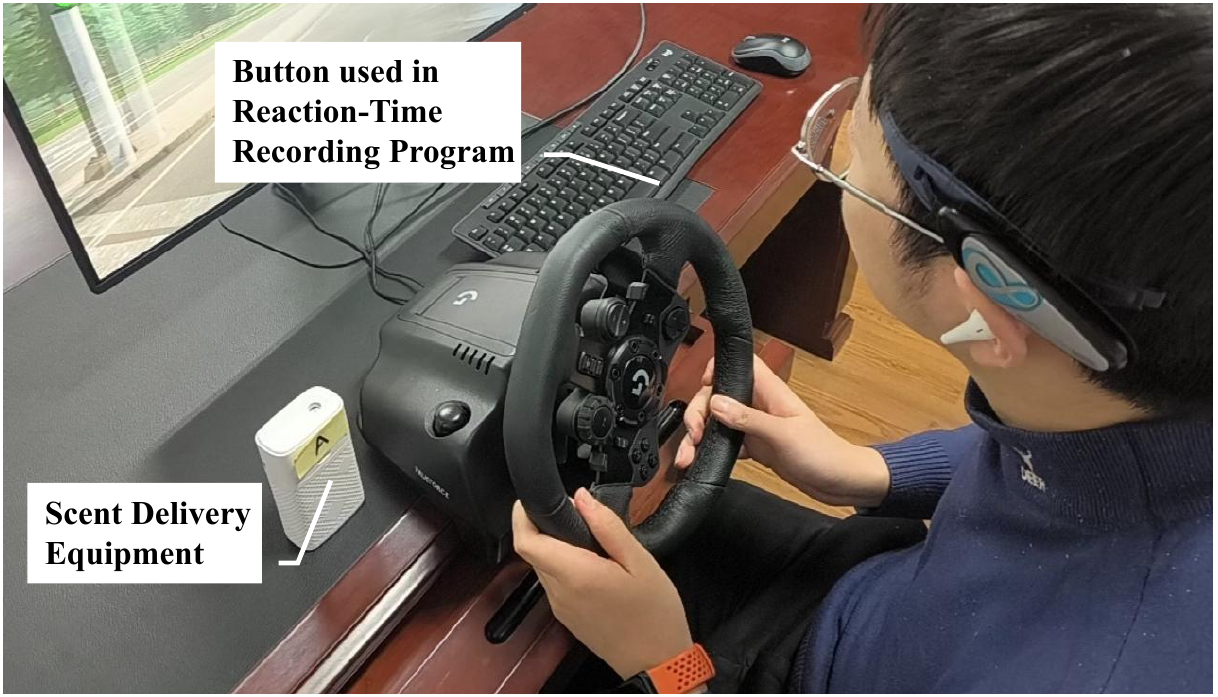}}
\subfloat[The scenes of the driving simulator]{
		\includegraphics[scale=0.35]{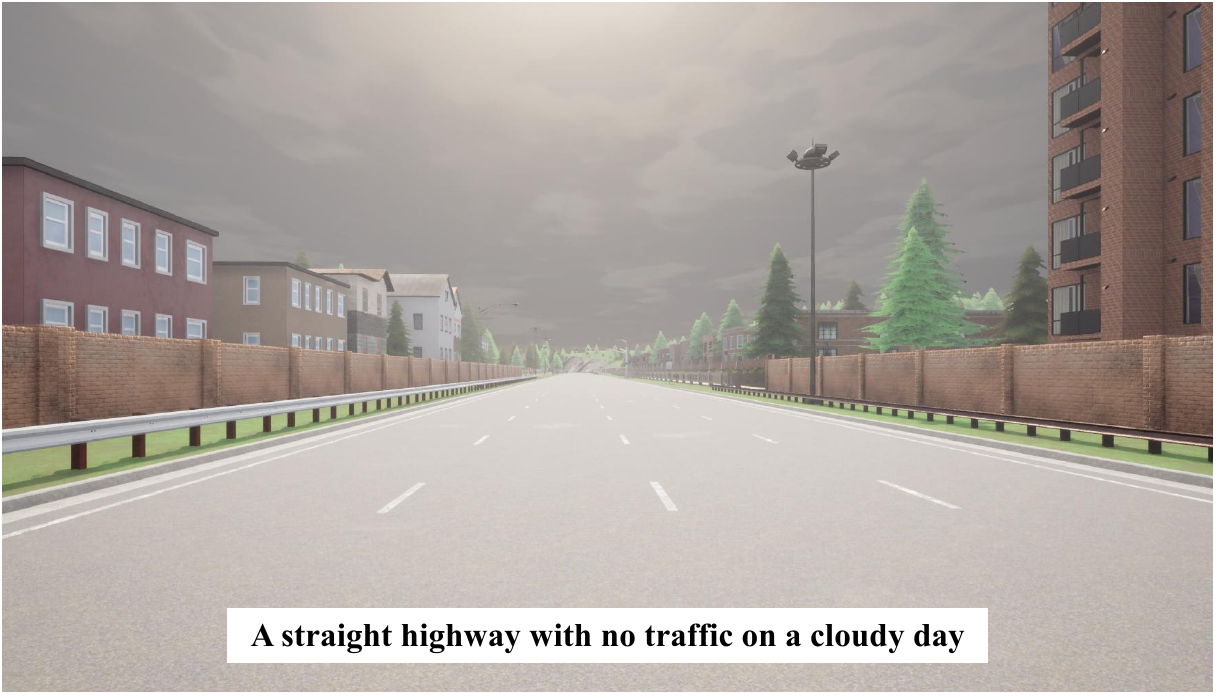}}
\caption{The on-site equipment for Study 2: Fig. (a) shows the position of the steering wheel, the buttons of the reaction-time recording program, and the scent release device. Fig. (b) shows the driving simulation scene used throughout Study 2. }
\label{fig:3}
\end{figure}

\subsection{Study 2:Exploring the effects of argy wormwood and tangerine peel on driving fatigue}

\subsubsection{Design}
This study followed a within-subjects experimental design of 1 (state: fatigued) × 3 (scents: tangerine peel, argy wormwood, water). It mainly measures the impact of TCM scents on three indicators: alertness, driving performance, and acceptance.

\footnotetext[1]{Generally, KSS is a 9-level scale. However, due to the limited number of participants in this experiment, for the convenience of data analysis, levels 2, 4, 6, and 8 were removed, and only levels 1, 3, 5, 7, and 9 were retained to form a 5-level scale. }

\begin{itemize}
    \item \textbf{Alertness:} \textit{Karolinska Sleepiness Scale (KSS)\cite{lee2020making}:} It is a subjective evaluation scale to measure the degree of sleepiness. 1 - 5 represents awake - sleepy\footnotemark[1].
    \item \textbf{Driving performance:}\textit{reaction-time:} Participants quickly press a button to record the reaction-time when they hear a beep.
    \item \textbf{Acceptance:} \textit{Questionnaire:} The questionnaire examines three indicators of scent acceptance: liking, comfort, and intensity\cite{dmitrenko2020caroma}. 1 - 5 represents low - high degree. \textit{Interview:} The interview examines whether participants can clearly smell the scent, the duration of the scent effectiveness and whether they are willing to choose this scent as a regulator for driving fatigue and why. 
\end{itemize}

\subsubsection{Setup}
\begin{itemize}
    \item Driving Simulator: Participants sit in the seat of the driving simulator equipped with a Logitech G29 Driving Force steering wheel and pedals(see Fig.~\ref{fig:3}(a)). The front main screen (27-inch, 60Hz refresh rate) shows the view outside the vehicle. The driving simulation scenario is built with Carla\cite{dosovitskiy2017carla}. The simulated scenario is a straight highway with no traffic on a cloudy day(see Fig.~\ref{fig:3}(b))\cite{huang2024chatbot}. A Python program is used to record the reaction-time.
    \item Scent Delivery Equipment: A Nimin aroma diffuser, which can be connected to the Mi Home APP to adjust the scent-emitting concentration and interval time. Three diffusers contain diluted essential oils (essential oil: alcohol: water = 1:6:3) (as in \cite{jiang2023study})of tangerine peel, argy wormwood, and water. The scent delivery equipment is located on the left side of the steering wheel.
\end{itemize}

\subsubsection{Procedure}

\begin{figure}
    \centering
    \includegraphics[width=1\linewidth]{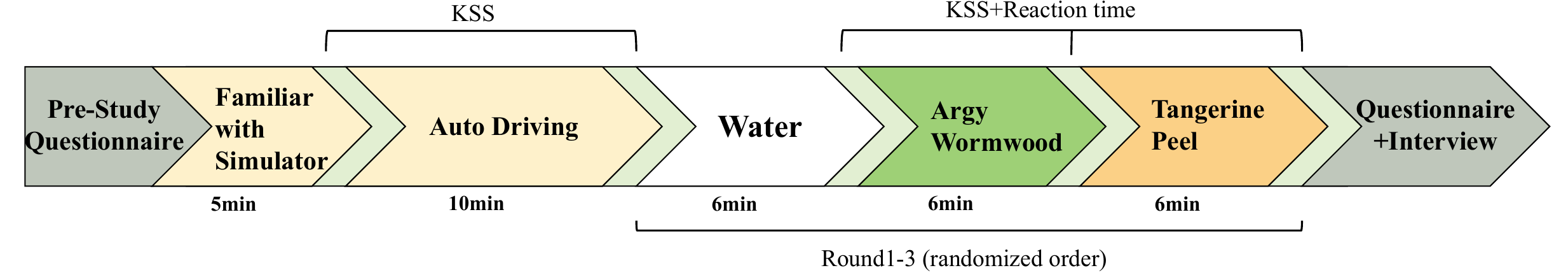}
    \caption{The timeline of the Study 2 procedure. Participants filled out a pre-questionnaire. Then, they drove autonomously for 5min to get used to the simulator. Next, they drove autonomously for 10min to induce fatigue. Then, 3 rounds of scent release experiments were conducted. The round order was randomized. Finally, participants filled out a scent-evaluation questionnaire and had an interview.}
    \label{fig:4}
\end{figure}

As in Fig. ~\ref{fig:4}, participants read, signed the consent form, and filled out a pre-questionnaire on basic info like gender and age. Then, they put on headphones, tested the volume and the reaction-time recording program, and drove autonomously for 5min in a simulated highway scene, hands on the wheel and feet on the brake to get familiar with the program and driving simulator requirements. Next, they used the KSS questionnaire to rate fatigue. Then, participants engaged in driving fatigue induction experiment for 10min. After that, they filled out the KSS again to confirm the effect. Subsequently, three rounds of scent release experiments were administered. Scents were released at 0, 3, and 6min for 30s in each round. Participants pressed a button after a beep at 6 min, with reaction-time recorded and KSS filled out. The round order was randomized by Latin-square design to mitigate
sequence effects\cite{fisher1970statistical}. Finally, participants filled out a scent-evaluation questionnaire, had an interview, and got 50RMB for participation. The whole experiment took about 70min. 

\subsubsection{Results}
\paragraph{Participants}
20 participants, 22-28 years of age (M = 24.7, SD = 1.45, 8 females).They were asked to be well-rested and avoid caffeine or alcohol before the study. Based on interview results, data from 3 participants who were unsuccessfully induced into driving fatigue or were unfocused during the experiment were excluded. 

\begin{figure}[!t]
\centering
\subfloat[KSS before and after driving fatigue induction experiment]{
		\includegraphics[scale=0.45]{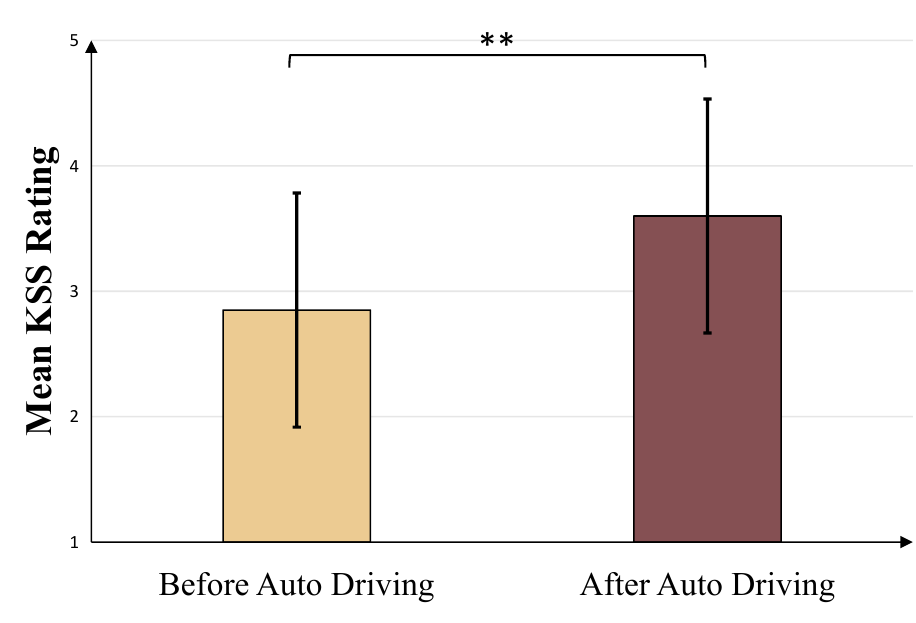}}
\subfloat[KSS after scent release experiments]{
		\includegraphics[scale=0.45]{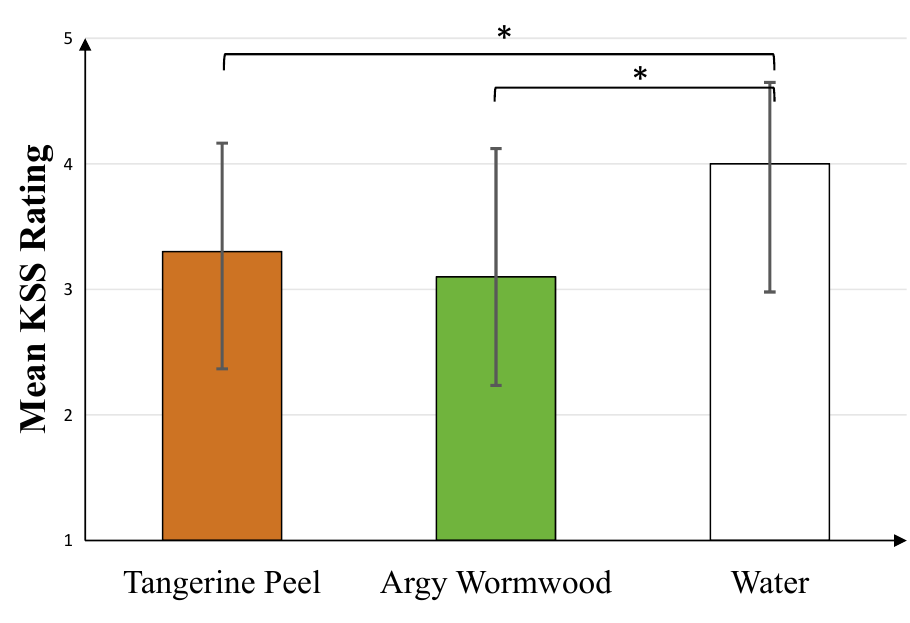}}
\\
\subfloat[Reaction-time after scent release experiments]{
		\includegraphics[scale=0.45]{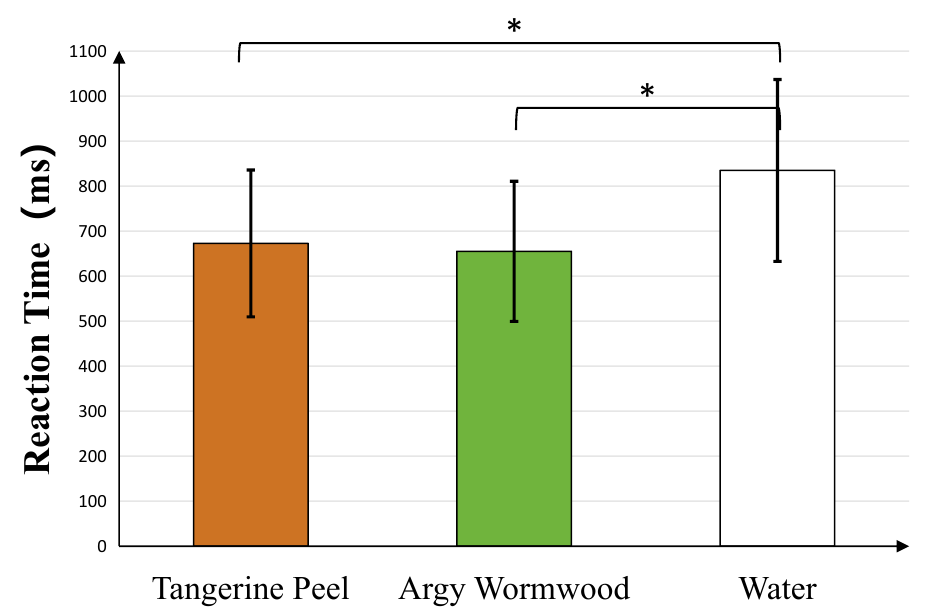}}
\subfloat[Acceptance of TCM scents]{
		\includegraphics[scale=0.45]{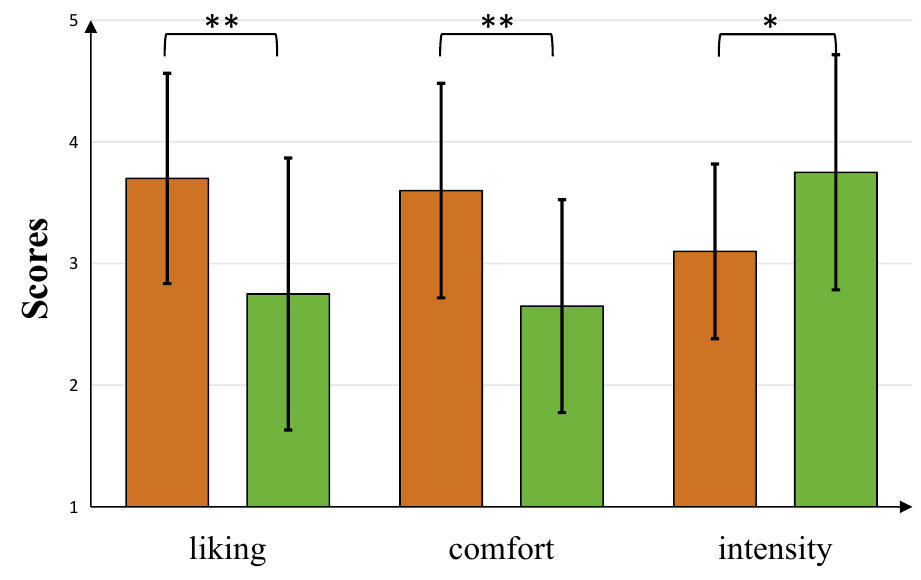}}
\caption{Results of Study 2. (* p < 0.05, ** p < 0.01, *** p < 0.001, all figures follow this annotation)}
\label{fig:5}
\end{figure}

\paragraph{Alertness before and after the driving fatigue induction experiment}
The Shapiro-Wilk test showed no variables were normally distributed. We used the Wilcoxon-Signed-Ranks test to compare the means between two variables. We compared KSS values before and after the driving fatigue induction experiment (see Fig. ~\ref{fig:5}(a)). The test showed the  KSS value (M = 2.8, SD = 0.93) before the experiment was significantly lower than the one (M = 3.6, SD = 0.94, Z = 2.762, p < 0.01) after the experiment. This means that the participants were awake at the start of the study and reported feeling fatigued after the driving fatigue induction experiment.

\paragraph{Alertness after scent release experiments}
A repeated-measures ANOVA on KSS scores showed scent types had a main effect on KSS ratings (F = 3.917, p < 0.05)(see Fig. ~\ref{fig:5}(b)), indicating scent types can affect drivers' subjective alertness scores. The Friedman test showed drivers' subjective alertness under argy wormwood and tangerine peel was higher than the baseline (water) (p < 0.05), yet there was no significant difference between argy wormwood and tangerine peel (p = 0.29). This means both TCM scents can improve drivers' subjective alertness. Argy wormwood's refreshing effect(M = 3.3, SD = 0.86) is slightly stronger than tangerine peel's(M = 3.1, SD = 1.02), but not significantly so. 

\paragraph{Driving performance after scent release experiments}
A repeated-measures ANOVA on reaction-time showed scent types had a main effect on driving performance (F = 3.95, p < 0.05)(see Fig. ~\ref{fig:5} (c)), indicating scent types can affect drivers' driving performance. The Friedman test showed drivers' driving performance under argy wormwood and tangerine peel was better than the baseline (water) (p < 0.05) , yet there was no significant difference between argy wormwood and tangerine peel (p = 0.85). This means both TCM scents can improve driving performance, with no significant difference between them.

\begin{figure}
    \centering
    \includegraphics[width=0.95\linewidth]{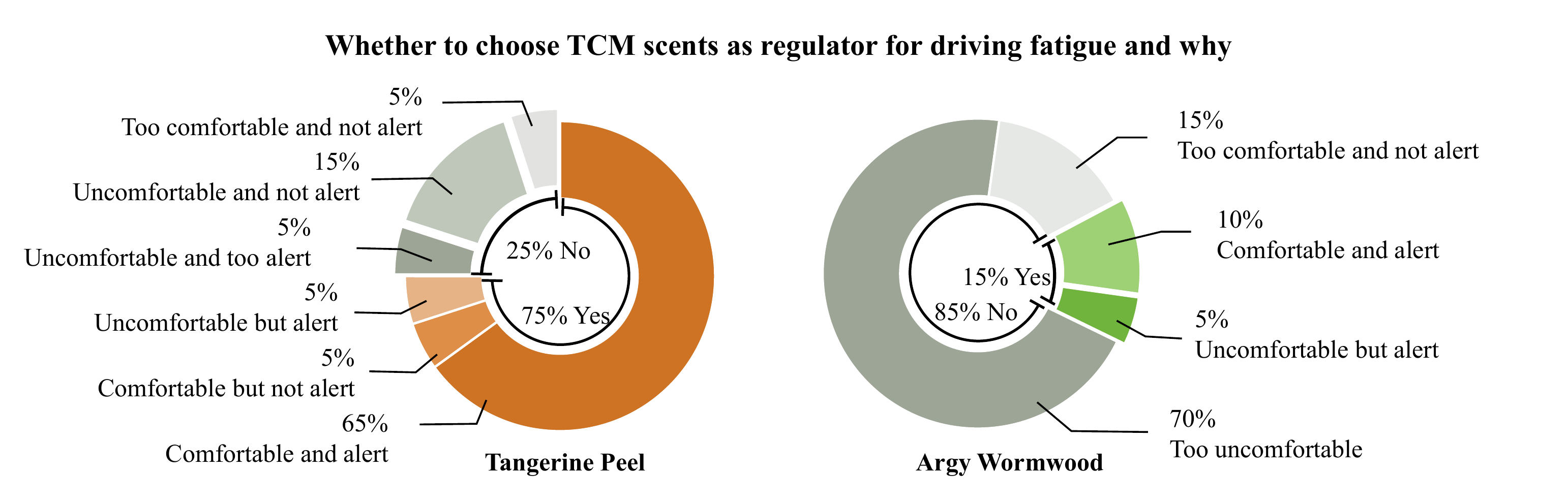}
    \caption{Interview results in Study 2: Most people are willing to use tangerine peel for driving fatigue regulation because it's comfortable and can help them stay alert. Most people are not willing to use argy wormwood because it's uncomfortable.}
    \label{fig:6}
\end{figure}

\paragraph{Acceptance of TCM scents}
A repeated-measures ANOVA on questionnaire data showed scent types had a main effect on liking (F = 8.99, p < 0.01), comfort (F = 11.60, p < 0.01), and intensity (F = 5.71, p < 0.05)(see Fig. ~\ref{fig:6}), indicating scent types can affect drivers' acceptance. Paired-sample tests showed argy wormwood's liking(p < 0.01)and comfort(p < 0.01) were significantly lower than tangerine peel's. But argy wormwood's intensity(p < 0.05) was significantly higher than tangerine peel's(see Fig. ~\ref{fig:5}(d)). Interview data  showed 75\% of participants would choose tangerine peel as a regulator for driving fatigue because it can mitigate driving fatigue and it is comfortable. Only 15\% would choose argy wormwood due to its discomfort.

\section{Discussion and Future Work}

\subsection{Discussion}
\begin{figure}
    \centering
    \includegraphics[width=0.95\linewidth]{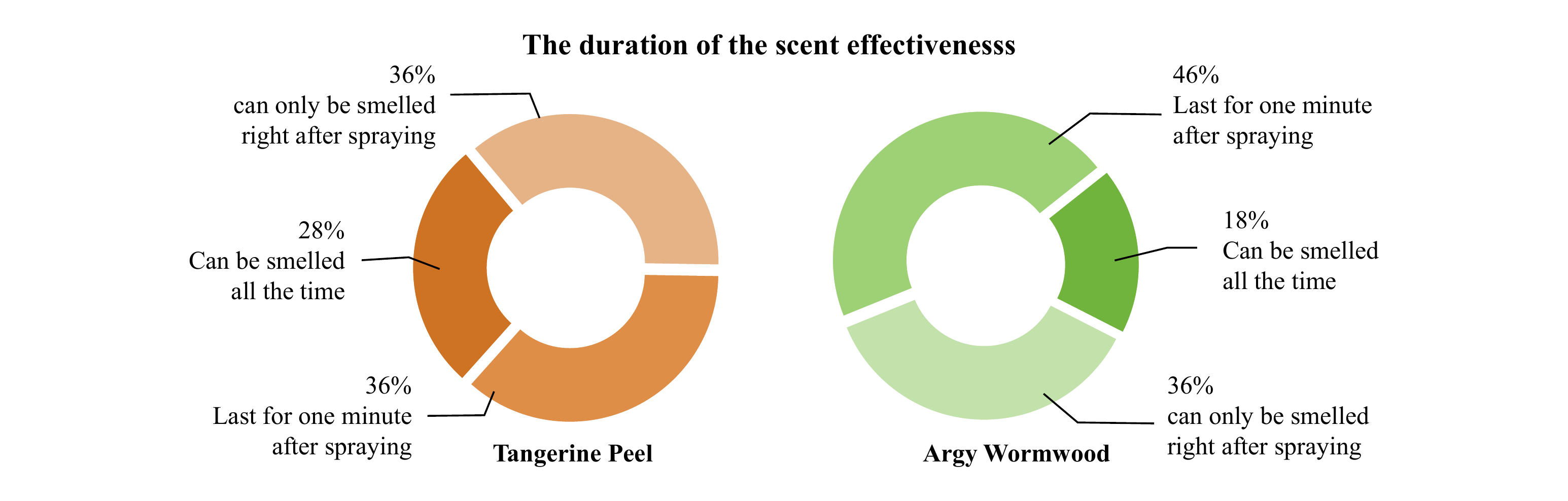}
    \caption{Feasibility of TCM scents as regulator for driving fatigue: During the experiment, only a few participants could smell the TCM scents all the time, indicating the scents' short-lasting time. }
    \label{fig:7}
\end{figure}

\subsubsection{Effect of TCM scents on driving fatigue} 
The two chosen TCM scents both have clear effects on mitigating driving fatigue. They can effectively boost drivers' alertness, reduce the fatigue induced by autonomous driving, and thereby enhance driving safety. Between the two, argy wormwood exhibits a slightly stronger effect than tangerine peel, although the difference is not substantial. This might be due to the narrow gap in arousal levels between argy wormwood and tangerine peel. Moreover, tangerine peel has a higher valence, which can provide a more enjoyable driving experience.

\subsubsection{Effect of TCM scents on driving performance}
 Upon exposure to the TCM scents, the participants exhibited a significant reduction in reaction-time, thereby indicating an improvement in driving performance. This outcome is advantageous for enhancing driving safety. Consistent with the findings related to alertness, no statistically significant difference was observed between the effects of argy wormwood and tangerine peel.

\subsubsection{Acceptance of TCM scents}
The acceptance of the two TCM scents differs significantly. Tangerine peel has a higher level of liking and comfort but a lower intensity, while argy wormwood has a lower level of liking and comfort but a higher intensity. Based on the interview results, tangerine peel is perceived as a more favorable regulator for driving fatigue, compared to argy wormwood. Therefore, tangerine peel may be more suitable for long-term application in driving fragrance due to its comforting and refreshing attributes. On the contrary, argy wormwood may be more appropriate as an alertness-enhancing scent during periods of fatigue. It can be activated either actively or passively when drivers experience severe drowsiness, as it possesses a notable refreshing effect, even though it is not as comforting.

\subsection{Future Work}
\subsubsection{Feasibility of TCM scents as regulator for driving fatigue}
 The duration of the scent effectiveness for TCM scents is relatively short, as illustrated in Fig.~\ref{fig:7}. Maybe it's because TCM essential oils are mainly used for massage or medicinal baths, not as specialized fragrance essential oils. Unlike regular fragrances, TCM essential oils lack top, middle, and base notes. Mixing multiple essential oils to make compound essential oils might prolong the duration.

\subsubsection{Auto-regulation system for mitigating driving fatigue by combining AI and TCM scents}
 AI can determine whether a driver is fatigued and the extent of their fatigue by analyzing both in-car and out-car information. It can use data from the dashboard camera to assess whether the driver has been engaged in prolonged monotonous driving, and it can utilize in-car cameras to recognize facial expressions\cite{zhu2022research} or upper body posture variations\cite{ansari2022automatic} to judge if the driver is experiencing fatigue. Based on the driver's condition and the characteristics of TCM scents, AI can generate appropriate adjustment strategies to mitigate driving fatigue.
\section{Conclusion}
The results of this study prove that the TCM scents can mitigate driving fatigue induced by autonomous driving, and enhance the drivers' alertness and driving performance. Among them, argy wormwood has a relatively high intensity and a better refreshing effect, but its liking and comfort level are relatively low. So it is more suitable as a short-term reminder scent. Tangerine peel has a relatively low intensity, but a higher liking and comfort level, making it more suitable as a in-car fragrance for long-term use. This study pioneers the exploration of the application of TCM scents in mitigating driving fatigue, offering invaluable insights for the future development of in-car systems. 

\bibliographystyle{ACM-Reference-Format}
\bibliography{ref}


\begin{thebibliography}{34}


\ifx \showCODEN    \undefined \def \showCODEN     #1{\unskip}     \fi
\ifx \showDOI      \undefined \def \showDOI       #1{#1}\fi
\ifx \showISBNx    \undefined \def \showISBNx     #1{\unskip}     \fi
\ifx \showISBNxiii \undefined \def \showISBNxiii  #1{\unskip}     \fi
\ifx \showISSN     \undefined \def \showISSN      #1{\unskip}     \fi
\ifx \showLCCN     \undefined \def \showLCCN      #1{\unskip}     \fi
\ifx \shownote     \undefined \def \shownote      #1{#1}          \fi
\ifx \showarticletitle \undefined \def \showarticletitle #1{#1}   \fi
\ifx \showURL      \undefined \def \showURL       {\relax}        \fi
\providecommand\bibfield[2]{#2}
\providecommand\bibinfo[2]{#2}
\providecommand\natexlab[1]{#1}
\providecommand\showeprint[2][]{arXiv:#2}

\bibitem[Jamson et~al\mbox{.}(2013)]%
        {jamson2013behavioural}
\bibfield{author}{\bibinfo{person}{A~Hamish Jamson}, \bibinfo{person}{Natasha Merat}, \bibinfo{person}{Oliver~MJ Carsten}, {and} \bibinfo{person}{Frank~CH Lai}.} \bibinfo{year}{2013}\natexlab{}.
\newblock \showarticletitle{Behavioural changes in drivers experiencing highly-automated vehicle control in varying traffic conditions}.
\newblock \bibinfo{journal}{\emph{Transportation research part C: emerging technologies}}  \bibinfo{volume}{30} (\bibinfo{year}{2013}), \bibinfo{pages}{116--125}.
\newblock
\urldef\tempurl%
\url{https://doi.org/10.1016/j.trc.2013.02.008}
\showDOI{\tempurl}


\bibitem[Eyben et~al\mbox{.}(2010)]%
        {eyben2010emotion}
\bibfield{author}{\bibinfo{person}{Florian Eyben}, \bibinfo{person}{Martin W{\"o}llmer}, \bibinfo{person}{Tony Poitschke}, \bibinfo{person}{Bj{\"o}rn Schuller}, \bibinfo{person}{Christoph Blaschke}, \bibinfo{person}{Berthold F{\"a}rber}, {and} \bibinfo{person}{Nhu Nguyen-Thien}.} \bibinfo{year}{2010}\natexlab{}.
\newblock \showarticletitle{Emotion on the road—necessity, acceptance, and feasibility of affective computing in the car}.
\newblock \bibinfo{journal}{\emph{Advances in Human-Computer Interaction}} \bibinfo{volume}{2010}, \bibinfo{number}{1} (\bibinfo{year}{2010}), \bibinfo{pages}{263593}.
\newblock
\urldef\tempurl%
\url{https://doi.org/10.1155/2010/263593}
\showDOI{\tempurl}


\bibitem[Li et~al\mbox{.}(2024)]%
        {li2024review}
\bibfield{author}{\bibinfo{person}{Wenbo Li}, \bibinfo{person}{Guofa Li}, \bibinfo{person}{Ruichen Tan}, \bibinfo{person}{Cong Wang}, \bibinfo{person}{Zemin Sun}, \bibinfo{person}{Ying Li}, \bibinfo{person}{Gang Guo}, \bibinfo{person}{Dongpu Cao}, {and} \bibinfo{person}{Keqiang Li}.} \bibinfo{year}{2024}\natexlab{}.
\newblock \showarticletitle{Review and Perspectives on Human Emotion for Connected Automated Vehicles}.
\newblock \bibinfo{journal}{\emph{Automotive Innovation}} \bibinfo{volume}{7}, \bibinfo{number}{1} (\bibinfo{year}{2024}), \bibinfo{pages}{4--44}.
\newblock
\urldef\tempurl%
\url{https://doi.org/10.1007/s42154-023-00270-z}
\showDOI{\tempurl}


\bibitem[Zhang et~al\mbox{.}(2016)]%
        {zhang2016traffic}
\bibfield{author}{\bibinfo{person}{Guangnan Zhang}, \bibinfo{person}{Kelvin~KW Yau}, \bibinfo{person}{Xun Zhang}, {and} \bibinfo{person}{Yanyan Li}.} \bibinfo{year}{2016}\natexlab{}.
\newblock \showarticletitle{Traffic accidents involving fatigue driving and their extent of casualties}.
\newblock \bibinfo{journal}{\emph{Accident Analysis \& Prevention}}  \bibinfo{volume}{87} (\bibinfo{year}{2016}), \bibinfo{pages}{34--42}.
\newblock
\urldef\tempurl%
\url{https://doi.org/10.1016/j.aap.2015.10.033}
\showDOI{\tempurl}


\bibitem[Li et~al\mbox{.}(2021)]%
        {li2021visual}
\bibfield{author}{\bibinfo{person}{Xiao-jun Li}, \bibinfo{person}{Jia-xin Ling}, {and} \bibinfo{person}{Yi Shen}.} \bibinfo{year}{2021}\natexlab{}.
\newblock \showarticletitle{Visual fatigue relief zone in an extra-long tunnel using virtual reality with wearable eeg-based devices}.
\newblock \bibinfo{journal}{\emph{Journal of Central South University}} \bibinfo{volume}{28}, \bibinfo{number}{12} (\bibinfo{year}{2021}), \bibinfo{pages}{3871--3881}.
\newblock
\urldef\tempurl%
\url{https://doi.org/10.1007/s11771-021-4882-8}
\showDOI{\tempurl}


\bibitem[Qin et~al\mbox{.}(2021)]%
        {qin2021characteristics}
\bibfield{author}{\bibinfo{person}{Pengcheng Qin}, \bibinfo{person}{Mingnian Wang}, \bibinfo{person}{Zhanwen Chen}, \bibinfo{person}{Guanfeng Yan}, \bibinfo{person}{Tao Yan}, \bibinfo{person}{Changling Han}, \bibinfo{person}{Yifan Bao}, {and} \bibinfo{person}{Xu Wang}.} \bibinfo{year}{2021}\natexlab{}.
\newblock \showarticletitle{Characteristics of driver fatigue and fatigue-relieving effect of special light belt in extra-long highway tunnel: A real-road driving study}.
\newblock \bibinfo{journal}{\emph{Tunnelling and Underground Space Technology}}  \bibinfo{volume}{114} (\bibinfo{year}{2021}), \bibinfo{pages}{103990}.
\newblock
\urldef\tempurl%
\url{https://doi.org/10.1016/j.tust.2021.103990}
\showDOI{\tempurl}


\bibitem[Hassib et~al\mbox{.}(2019)]%
        {hassib2019detecting}
\bibfield{author}{\bibinfo{person}{Mariam Hassib}, \bibinfo{person}{Michael Braun}, \bibinfo{person}{Bastian Pfleging}, {and} \bibinfo{person}{Florian Alt}.} \bibinfo{year}{2019}\natexlab{}.
\newblock \showarticletitle{Detecting and influencing driver emotions using psycho-physiological sensors and ambient light}. In \bibinfo{booktitle}{\emph{IFIP Conference on Human-Computer Interaction}}. Springer, \bibinfo{pages}{721--742}.
\newblock
\urldef\tempurl%
\url{https://doi.org/10.1007/978-3-030-29381-9_43}
\showDOI{\tempurl}


\bibitem[Orsini et~al\mbox{.}(2024)]%
        {orsini2024music}
\bibfield{author}{\bibinfo{person}{Federico Orsini}, \bibinfo{person}{Andrea Baldassa}, \bibinfo{person}{Massimo Grassi}, \bibinfo{person}{Nicola Cellini}, {and} \bibinfo{person}{Riccardo Rossi}.} \bibinfo{year}{2024}\natexlab{}.
\newblock \showarticletitle{Music as a countermeasure to fatigue: a driving simulator study}.
\newblock \bibinfo{journal}{\emph{Transportation research part F: traffic psychology and behaviour}}  \bibinfo{volume}{103} (\bibinfo{year}{2024}), \bibinfo{pages}{290--305}.
\newblock
\urldef\tempurl%
\url{https://doi.org/10.1016/j.trf.2024.04.016}
\showDOI{\tempurl}


\bibitem[Trumbo et~al\mbox{.}(2017)]%
        {trumbo2017name}
\bibfield{author}{\bibinfo{person}{Michael~C Trumbo}, \bibinfo{person}{Aaron~P Jones}, \bibinfo{person}{Charles~SH Robinson}, \bibinfo{person}{Kerstan Cole}, {and} \bibinfo{person}{James~D Morrow}.} \bibinfo{year}{2017}\natexlab{}.
\newblock \showarticletitle{Name that tune: Mitigation of driver fatigue via a song naming game}.
\newblock \bibinfo{journal}{\emph{Accident Analysis \& Prevention}}  \bibinfo{volume}{108} (\bibinfo{year}{2017}), \bibinfo{pages}{275--284}.
\newblock
\urldef\tempurl%
\url{https://doi.org/10.1016/j.aap.2017.09.002}
\showDOI{\tempurl}


\bibitem[Dahlman et~al\mbox{.}(2024)]%
        {dahlman2024vehicle}
\bibfield{author}{\bibinfo{person}{Anna~Sj{\"o}rs Dahlman}, \bibinfo{person}{Mikael~Ljung Aust}, \bibinfo{person}{Yaniv Mama}, \bibinfo{person}{Dan Hasson}, {and} \bibinfo{person}{Anna Anund}.} \bibinfo{year}{2024}\natexlab{}.
\newblock \showarticletitle{In-vehicle fragrance administration as a countermeasure for driver fatigue}.
\newblock \bibinfo{journal}{\emph{Accident Analysis \& Prevention}}  \bibinfo{volume}{195} (\bibinfo{year}{2024}), \bibinfo{pages}{107429}.
\newblock
\urldef\tempurl%
\url{https://doi.org/10.1016/j.aap.2023.107429}
\showDOI{\tempurl}


\bibitem[Jiang et~al\mbox{.}(2023)]%
        {jiang2023study}
\bibfield{author}{\bibinfo{person}{Xinyue Jiang}, \bibinfo{person}{Kanesan Muthusamy}, {and} \bibinfo{person}{Nagesparan Ainarappan}.} \bibinfo{year}{2023}\natexlab{}.
\newblock \showarticletitle{Study on Peppermint Scent Interventions for Fatigue Driving}. In \bibinfo{booktitle}{\emph{2023 IEEE European Technology and Engineering Management Summit (E-TEMS)}}. IEEE, \bibinfo{pages}{80--84}.
\newblock
\urldef\tempurl%
\url{https://doi.org/10.1109/E-TEMS57541.2023.10424578}
\showDOI{\tempurl}


\bibitem[Nan~Lv et~al\mbox{.}(2013)]%
        {nan2013aromatherapy}
\bibfield{author}{\bibinfo{person}{Xiao Nan~Lv}, \bibinfo{person}{Zhu Jun~Liu}, \bibinfo{person}{Huan Jing~Zhang}, {and} \bibinfo{person}{Chi~Meng Tzeng}.} \bibinfo{year}{2013}\natexlab{}.
\newblock \showarticletitle{Aromatherapy and the central nerve system (CNS): therapeutic mechanism and its associated genes}.
\newblock \bibinfo{journal}{\emph{Current drug targets}} \bibinfo{volume}{14}, \bibinfo{number}{8} (\bibinfo{year}{2013}), \bibinfo{pages}{872--879}.
\newblock


\bibitem[Giannenas et~al\mbox{.}(2020)]%
        {giannenas2020history}
\bibfield{author}{\bibinfo{person}{Ilias Giannenas}, \bibinfo{person}{E Sidiropoulou}, \bibinfo{person}{Eleftherios Bonos}, \bibinfo{person}{E Christaki}, {and} \bibinfo{person}{P Florou-Paneri}.} \bibinfo{year}{2020}\natexlab{}.
\newblock \showarticletitle{The history of herbs, medicinal and aromatic plants, and their extracts: Past, current situation and future perspectives}.
\newblock In \bibinfo{booktitle}{\emph{Feed additives}}. \bibinfo{publisher}{Elsevier}, \bibinfo{pages}{1--18}.
\newblock
\urldef\tempurl%
\url{https://doi.org/10.1016/B978-0-12-814700-9.00001-7}
\showDOI{\tempurl}


\bibitem[Dmitrenko et~al\mbox{.}(2020)]%
        {dmitrenko2020caroma}
\bibfield{author}{\bibinfo{person}{Dmitrijs Dmitrenko}, \bibinfo{person}{Emanuela Maggioni}, \bibinfo{person}{Giada Brianza}, \bibinfo{person}{Brittany~E Holthausen}, \bibinfo{person}{Bruce~N Walker}, {and} \bibinfo{person}{Marianna Obrist}.} \bibinfo{year}{2020}\natexlab{}.
\newblock \showarticletitle{Caroma therapy: pleasant scents promote safer driving, better mood, and improved well-being in angry drivers}. In \bibinfo{booktitle}{\emph{Proceedings of the 2020 chi conference on human factors in computing systems}}. \bibinfo{pages}{1--13}.
\newblock
\urldef\tempurl%
\url{https://doi.org/10.1145/3313831.3376176}
\showDOI{\tempurl}


\bibitem[Guo et~al\mbox{.}(2024)]%
        {guo2024could}
\bibfield{author}{\bibinfo{person}{Haofei Guo}, \bibinfo{person}{Jinxian Weng}, \bibinfo{person}{Kun Shi}, {and} \bibinfo{person}{Lei Wang}.} \bibinfo{year}{2024}\natexlab{}.
\newblock \showarticletitle{Could music reduce driver fatigue? An investigation on music effects in various weather conditions}.
\newblock \bibinfo{journal}{\emph{Journal of Transportation Safety \& Security}} (\bibinfo{year}{2024}), \bibinfo{pages}{1--22}.
\newblock
\urldef\tempurl%
\url{https://doi.org/10.1080/19439962.2024.2415959}
\showDOI{\tempurl}


\bibitem[Alaoui-Isma{\"\i}li et~al\mbox{.}(1997)]%
        {alaoui1997basic}
\bibfield{author}{\bibinfo{person}{Ouafe Alaoui-Isma{\"\i}li}, \bibinfo{person}{O Robin}, \bibinfo{person}{H Rada}, \bibinfo{person}{Andr{\'e} Dittmar}, {and} \bibinfo{person}{Evelyne Vernet-Maury}.} \bibinfo{year}{1997}\natexlab{}.
\newblock \showarticletitle{Basic emotions evoked by odorants: comparison between autonomic responses and self-evaluation}.
\newblock \bibinfo{journal}{\emph{Physiology \& behavior}} \bibinfo{volume}{62}, \bibinfo{number}{4} (\bibinfo{year}{1997}), \bibinfo{pages}{713--720}.
\newblock
\urldef\tempurl%
\url{https://doi.org/10.1016/S0031-9384(97)90016-0}
\showDOI{\tempurl}


\bibitem[Jiang et~al\mbox{.}(2024)]%
        {jiang2024scented}
\bibfield{author}{\bibinfo{person}{Xinyue Jiang}, \bibinfo{person}{Kanesan Muthusamy}, \bibinfo{person}{Jian Chen}, {and} \bibinfo{person}{Xueliang Fang}.} \bibinfo{year}{2024}\natexlab{}.
\newblock \showarticletitle{Scented Solutions: Examining the Efficacy of Scent Interventions in Mitigating Driving Fatigue}.
\newblock \bibinfo{journal}{\emph{Sensors}} \bibinfo{volume}{24}, \bibinfo{number}{8} (\bibinfo{year}{2024}), \bibinfo{pages}{2384}.
\newblock
\urldef\tempurl%
\url{https://doi.org/10.3390/s24082384}
\showDOI{\tempurl}


\bibitem[Yoshida et~al\mbox{.}(2011)]%
        {yoshida2011study}
\bibfield{author}{\bibinfo{person}{Mariko Yoshida}, \bibinfo{person}{Chie Kato}, \bibinfo{person}{Mikiko Kawasumi}, \bibinfo{person}{Hatsuo Yamasaki}, \bibinfo{person}{Shin Yamamoto}, \bibinfo{person}{Tomoaki Nakano}, {and} \bibinfo{person}{Muneo Yamada}.} \bibinfo{year}{2011}\natexlab{}.
\newblock \showarticletitle{Study on Stimulation Effects for Driver Based on Fragrance Presentation}. In \bibinfo{booktitle}{\emph{MVA}}. \bibinfo{pages}{332--335}.
\newblock


\bibitem[Yunjun(2013)]%
        {yunjun2013perfume}
\bibfield{author}{\bibinfo{person}{Chen Yunjun}.} \bibinfo{year}{2013}\natexlab{}.
\newblock \showarticletitle{The Perfume Culture of China and Taiwan: A Personal Report. 1}.
\newblock \bibinfo{journal}{\emph{Journal of the Royal Asiatic Society}} \bibinfo{volume}{23}, \bibinfo{number}{1} (\bibinfo{year}{2013}), \bibinfo{pages}{127--130}.
\newblock
\urldef\tempurl%
\url{https://doi.org/10.1017/S1356186313000059}
\showDOI{\tempurl}


\bibitem[Farrar and Farrar(2020)]%
        {farrar2020clinical}
\bibfield{author}{\bibinfo{person}{Ashley~J Farrar} {and} \bibinfo{person}{Francisca~C Farrar}.} \bibinfo{year}{2020}\natexlab{}.
\newblock \showarticletitle{Clinical aromatherapy}.
\newblock \bibinfo{journal}{\emph{The Nursing Clinics of North America}} \bibinfo{volume}{55}, \bibinfo{number}{4} (\bibinfo{year}{2020}), \bibinfo{pages}{489}.
\newblock


\bibitem[McCaffrey and Fowler(2003)]%
        {mccaffrey2003qigong}
\bibfield{author}{\bibinfo{person}{Ruth McCaffrey} {and} \bibinfo{person}{Nancy~L Fowler}.} \bibinfo{year}{2003}\natexlab{}.
\newblock \showarticletitle{Qigong practice: a pathway to health and healing}.
\newblock \bibinfo{journal}{\emph{Holistic nursing practice}} \bibinfo{volume}{17}, \bibinfo{number}{2} (\bibinfo{year}{2003}), \bibinfo{pages}{110--116}.
\newblock


\bibitem[Zhou et~al\mbox{.}(2021)]%
        {zhou2021conceptualization}
\bibfield{author}{\bibinfo{person}{Pin Zhou}, \bibinfo{person}{Hugo Critchley}, \bibinfo{person}{Sarah Garfinkel}, {and} \bibinfo{person}{Ya Gao}.} \bibinfo{year}{2021}\natexlab{}.
\newblock \showarticletitle{The conceptualization of emotions across cultures: a model based on interoceptive neuroscience}.
\newblock \bibinfo{journal}{\emph{Neuroscience \& Biobehavioral Reviews}}  \bibinfo{volume}{125} (\bibinfo{year}{2021}), \bibinfo{pages}{314--327}.
\newblock
\urldef\tempurl%
\url{https://doi.org/10.1016/j.neubiorev.2021.02.023.}
\showDOI{\tempurl}


\bibitem[Li et~al\mbox{.}(2020)]%
        {li2020traditional}
\bibfield{author}{\bibinfo{person}{Chan Li}, \bibinfo{person}{Junying Huang}, \bibinfo{person}{Yung-Chi Cheng}, {and} \bibinfo{person}{Yuan-Wei Zhang}.} \bibinfo{year}{2020}\natexlab{}.
\newblock \showarticletitle{Traditional Chinese medicine in depression treatment: from molecules to systems}.
\newblock \bibinfo{journal}{\emph{Frontiers in Pharmacology}}  \bibinfo{volume}{11} (\bibinfo{year}{2020}), \bibinfo{pages}{586}.
\newblock
\urldef\tempurl%
\url{https://doi.org/10.3389/fphar.2020.00586}
\showDOI{\tempurl}


\bibitem[Zhang and Cheng(2019)]%
        {zhang2019challenge}
\bibfield{author}{\bibinfo{person}{Yuan-Wei Zhang} {and} \bibinfo{person}{Yung-Chi Cheng}.} \bibinfo{year}{2019}\natexlab{}.
\newblock \showarticletitle{Challenge and prospect of traditional Chinese medicine in depression treatment}.
\newblock \bibinfo{journal}{\emph{Frontiers in neuroscience}}  \bibinfo{volume}{13} (\bibinfo{year}{2019}), \bibinfo{pages}{190}.
\newblock
\urldef\tempurl%
\url{https://doi.org/10.3389/fnins.2019.00190}
\showDOI{\tempurl}


\bibitem[Sun et~al\mbox{.}(2022)]%
        {sun2022dissecting}
\bibfield{author}{\bibinfo{person}{Yilu Sun}, \bibinfo{person}{Jia Zhao}, {and} \bibinfo{person}{Jianhui Rong}.} \bibinfo{year}{2022}\natexlab{}.
\newblock \showarticletitle{Dissecting the molecular mechanisms underlying the antidepressant activities of herbal medicines through the comprehensive review of the recent literatures}.
\newblock \bibinfo{journal}{\emph{Frontiers in Psychiatry}}  \bibinfo{volume}{13} (\bibinfo{year}{2022}), \bibinfo{pages}{1054726}.
\newblock
\urldef\tempurl%
\url{https://doi.org/10.3389/fpsyt.2022.1054726}
\showDOI{\tempurl}


\bibitem[Commission(2020)]%
        {2020Pharmacopoeia}
\bibfield{author}{\bibinfo{person}{Chinese~Pharmacopoeia Commission}.} \bibinfo{year}{2020}\natexlab{}.
\newblock \bibinfo{booktitle}{\emph{Pharmacopoeia of the People's Republic of China 2020}}.
\newblock \bibinfo{publisher}{China Medical Science Press}.
\newblock


\bibitem[Wilson and Brewster(2017)]%
        {wilson2017multi}
\bibfield{author}{\bibinfo{person}{Graham Wilson} {and} \bibinfo{person}{Stephen~A Brewster}.} \bibinfo{year}{2017}\natexlab{}.
\newblock \showarticletitle{Multi-moji: Combining thermal, vibrotactile \& visual stimuli to expand the affective range of feedback}. In \bibinfo{booktitle}{\emph{Proceedings of the 2017 CHI Conference on Human Factors in Computing Systems}}. \bibinfo{pages}{1743--1755}.
\newblock
\urldef\tempurl%
\url{https://doi.org/10.1145/3025453.3025614}
\showDOI{\tempurl}


\bibitem[Bradley and Lang(1994)]%
        {bradley1994measuring}
\bibfield{author}{\bibinfo{person}{Margaret~M Bradley} {and} \bibinfo{person}{Peter~J Lang}.} \bibinfo{year}{1994}\natexlab{}.
\newblock \showarticletitle{Measuring emotion: the self-assessment manikin and the semantic differential}.
\newblock \bibinfo{journal}{\emph{Journal of behavior therapy and experimental psychiatry}} \bibinfo{volume}{25}, \bibinfo{number}{1} (\bibinfo{year}{1994}), \bibinfo{pages}{49--59}.
\newblock
\urldef\tempurl%
\url{https://doi.org/10.1016/0005-7916(94)90063-9}
\showDOI{\tempurl}


\bibitem[Lee et~al\mbox{.}(2020)]%
        {lee2020making}
\bibfield{author}{\bibinfo{person}{Jieun Lee}, \bibinfo{person}{Toshiaki Hirano}, {and} \bibinfo{person}{Makoto Itoh}.} \bibinfo{year}{2020}\natexlab{}.
\newblock \showarticletitle{Making Passenger Conversation in Partial Driving Automation: Effects of Relationship Between Driver and Passenger on Driver Fatigue and Driving Performance}. In \bibinfo{booktitle}{\emph{2020 IEEE International Conference on Systems, Man, and Cybernetics (SMC)}}. IEEE, \bibinfo{pages}{1718--1722}.
\newblock
\urldef\tempurl%
\url{https://doi.org/10.1109/smc42975.2020.9283363}
\showDOI{\tempurl}


\bibitem[Dosovitskiy et~al\mbox{.}(2017)]%
        {dosovitskiy2017carla}
\bibfield{author}{\bibinfo{person}{Alexey Dosovitskiy}, \bibinfo{person}{German Ros}, \bibinfo{person}{Felipe Codevilla}, \bibinfo{person}{Antonio Lopez}, {and} \bibinfo{person}{Vladlen Koltun}.} \bibinfo{year}{2017}\natexlab{}.
\newblock \showarticletitle{CARLA: An open urban driving simulator}. In \bibinfo{booktitle}{\emph{Conference on robot learning}}. PMLR, \bibinfo{pages}{1--16}.
\newblock


\bibitem[Huang et~al\mbox{.}(2024)]%
        {huang2024chatbot}
\bibfield{author}{\bibinfo{person}{Shaoshuai Huang}, \bibinfo{person}{Xuandong Zhao}, \bibinfo{person}{Dapeng Wei}, \bibinfo{person}{Xinheng Song}, {and} \bibinfo{person}{Yuanbo Sun}.} \bibinfo{year}{2024}\natexlab{}.
\newblock \showarticletitle{Chatbot and Fatigued Driver: Exploring the Use of LLM-Based Voice Assistants for Driving Fatigue}. In \bibinfo{booktitle}{\emph{Extended Abstracts of the CHI Conference on Human Factors in Computing Systems}}. \bibinfo{pages}{1--8}.
\newblock
\urldef\tempurl%
\url{https://doi.org/10.1145/3613905.3651031}
\showDOI{\tempurl}


\bibitem[Fisher(1970)]%
        {fisher1970statistical}
\bibfield{author}{\bibinfo{person}{Ronald~Aylmer Fisher}.} \bibinfo{year}{1970}\natexlab{}.
\newblock \showarticletitle{Statistical methods for research workers}.
\newblock In \bibinfo{booktitle}{\emph{Breakthroughs in statistics: Methodology and distribution}}. \bibinfo{publisher}{Springer}, \bibinfo{pages}{66--70}.
\newblock
\urldef\tempurl%
\url{https://doi.org/10.1007/bf01603396}
\showDOI{\tempurl}


\bibitem[Zhu et~al\mbox{.}(2022)]%
        {zhu2022research}
\bibfield{author}{\bibinfo{person}{Tianjun Zhu}, \bibinfo{person}{Chuang Zhang}, \bibinfo{person}{Tunglung Wu}, \bibinfo{person}{Zhuang Ouyang}, \bibinfo{person}{Houzhi Li}, \bibinfo{person}{Xiaoxiang Na}, \bibinfo{person}{Jianguo Liang}, {and} \bibinfo{person}{Weihao Li}.} \bibinfo{year}{2022}\natexlab{}.
\newblock \showarticletitle{Research on a real-time driver fatigue detection algorithm based on facial video sequences}.
\newblock \bibinfo{journal}{\emph{Applied Sciences}} \bibinfo{volume}{12}, \bibinfo{number}{4} (\bibinfo{year}{2022}), \bibinfo{pages}{2224}.
\newblock
\urldef\tempurl%
\url{https://doi.org/10.3390/app12042224}
\showDOI{\tempurl}


\bibitem[Ansari et~al\mbox{.}(2022)]%
        {ansari2022automatic}
\bibfield{author}{\bibinfo{person}{Shahzeb Ansari}, \bibinfo{person}{Haiping Du}, \bibinfo{person}{Fazel Naghdy}, {and} \bibinfo{person}{David Stirling}.} \bibinfo{year}{2022}\natexlab{}.
\newblock \showarticletitle{Automatic driver cognitive fatigue detection based on upper body posture variations}.
\newblock \bibinfo{journal}{\emph{Expert Systems with Applications}}  \bibinfo{volume}{203} (\bibinfo{year}{2022}), \bibinfo{pages}{117568}.
\newblock
\urldef\tempurl%
\url{https://doi.org/10.1016/j.eswa.2022.117568}
\showDOI{\tempurl}


\end{thebibliography}

\appendix
\section{Pre-study}
In order to explore the proper duration of the driving fatigue induction experiment and the scent concentration of the scent release experiment in Study 2, we designed a Pre-study to conduct the exploration. 
\subsection{Design}
This study adopts a within-subjects design, and it mainly consists of three steps: 
\begin{enumerate}
    \item Get familiar with the driving simulator.
    \item Record the time when participants reach a state of general fatigue (Level 7 on the KSS, described as: Sleepy, but no effort to keep awake) and extreme fatigue(Level 9 on the KSS, described as: Very sleepy, great effort to keep awake, fighting sleep).
    \item Conduct three rounds of scent release experiments. In each round, argy wormwood at different concentrations will be used. Finally, participants select the round with the most comfortable concentration. The comfort criterion is that the scent can be clearly smelled and not too pungent. 
\end{enumerate}

\subsection{Setup}
The experimental equipment and the on-site settings are kept consistent with those in Study 2.

\subsection{Procedure}
\begin{figure}
    \centering
    \includegraphics[width=1\linewidth]{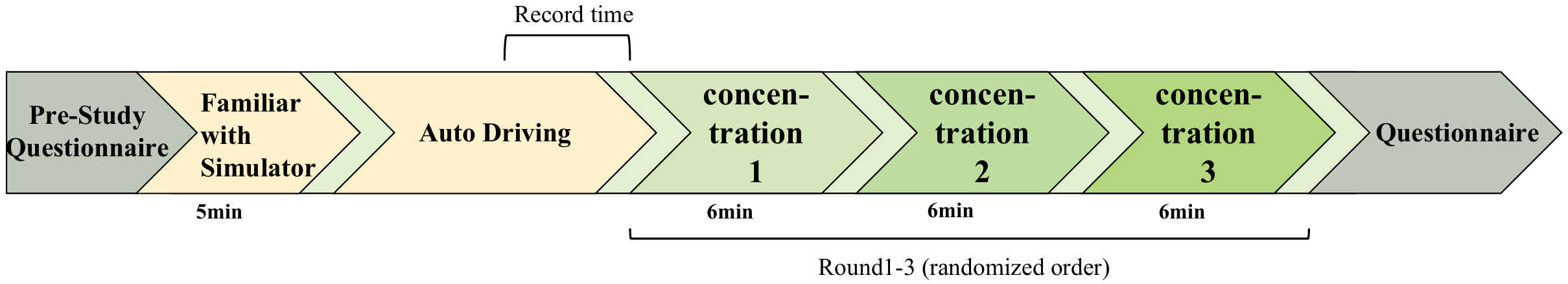}
    \caption{The timeline of the pre-study procedure. Participants filled out a pre-questionnaire. Then, they drove autonomously for 5min to get used to the simulator. Next, they drove autonomously and recorded the time when they were induced to driving fatigue. Then, 3 rounds of scent release experiments were conducted. The round order was randomized. Finally, participants filled out a questionnaire to choose the most comfortable round.}
    \label{fig:app-1}
\end{figure}

As in Fig. ~\ref{fig:app-1}, the participants read, signed the consent form and completed a pre-questionnaire on basic information such as gender and age. Then, they drove autonomously for 5min in a simulated highway scene, hands on the wheel and feet on the brake to get familiar with the driving simulator requirements. Next, they engaged in a driving fatigue induction experiment and recorded the time when they reached a state of general drowsiness and extreme drowsiness. The simulator scenario was the same as that in Study 2. The maximum duration of this experiment is 25min. If the participants still have not reached a state of extreme fatigue, the experiment will be stopped, and the data will be marked as "failure to induce fatigue".  After that, three rounds of scent release experiments were administered. The concentration of scent in each round was controlled by regulating the scent release time each time. The release times for the three rounds were 8s every three minutes, 16s every three minutes, and 30s every three minutes respectively. The round order was randomized by Latin-square design to mitigate sequence effects\cite{fisher1970statistical}. Finally, the participants were required to select the round with the most comfortable scent concentration.

\subsection{Results}
\paragraph{Participants}7 participants, 23-25 years old (M = 24.1, SD = 1.21, 3 females), volunteered for this study.

\paragraph{The proper duration of the driving fatigue induction experiment}Since the data of one participant was marked as "failure to induce fatigue", we only calculated the mean and variance of six groups of data. The calculation shows that the average time for participants to reach a state of general fatigue is 3.5min (SD = 1.87), and the average time to reach a state of extreme fatigue is 7.5min (SD = 1.38). In order to ensure that the participants are induced to experience driving fatigue, we selected 10min as the duration of the driving fatigue induction experiment in Study 2.

\paragraph{The proper concentration of the scent release experiment}Among the seven experiment participants, two chose the lowest concentration because they were allergic to the smell of traditional Chinese medicine, and a high concentration might irritate them and cause sneezing. The remaining five participants chose the highest concentration because of the short duration of scent effectiveness. Therefore, in the formal experiment, we screened out the participants who were allergic to the TCM scents and selected the highest concentration as the experimental concentration for Study 2. Due to the limitations of the equipment, we were unable to explore the impact of higher concentrations on the participants. This may be further studied in future experiments. 

\end{document}